\documentclass[prd,a4paper,byrevtex]{revtex4}
\begin{document}

\title{Comment on `Glueball spectrum from a potential model'}
\author{Fabian \surname{Brau}}
\email[E-mail: ]{fabian.brau@umh.ac.be}
\author{Claude \surname{Semay}}
\email[E-mail: ]{claude.semay@umh.ac.be}
\affiliation{Service de Physique G\'en\'erale et de Physique des
Particules El\'ementaires, Groupe de Physique Nucl\'eaire Th\'eorique,
Universit\'e de Mons-Hainaut, Place du Parc 20, B-7000 Mons, Belgium}
\date{\today}

\begin{abstract}
In a recent article, W.-S. Hou and G.-G. Wong [Phys. Rev. D {\bf 67},
034003 (2003)] have investigated the spectrum of two-gluon glueballs
below 3 GeV in a potential model with a dynamical gluon mass. We point
out that, among the 18 states calculated by the authors, only three are
physical. The other states either are spurious or possess a finite mass
only due to an arbitrary restriction of the variational parameter.
\end{abstract}



\maketitle

In Ref.~\cite{hou03}, the authors have studied the bound states of
confined gluons, known as glueballs, with a potential model  proposed
two decades ago by Cornwall and Soni \cite{cor83,hou84}. The motivation
was to cast some light on the glueballs properties with a simple model,
which allows the calculation of the spectrum of glueballs composed of
two, three or more gluons. This article \cite{hou03} was an extension of
a previous article \cite{hou01} where only a vanishing angular momentum
was considered.

The potential model used by Hou and Wong is potentially interesting but
the interaction between two gluons used to calculate the glueball
spectrum have two kind of singularities. The potential between two
gluons reads
\begin{eqnarray}
\label{eq1}
V_{2g}(r)=&-&\lambda\left\{\left[\frac{1}{4}+\frac{1}{3} {\bf S}^2+\frac
{3}{2m^2}({\bf L}\cdot{\bf S}) \frac{1}{r}\frac{\partial}{\partial r}-
\frac{1}{2m^2}\left(({\bf S}\cdot{\bf \nabla})^2-\frac{1}{3}{\bf S}^2
\nabla^2\right)\right]\frac{e^{-mr}}{r}+\left(1-\frac{5}{6}{\bf S}^2
\right)\frac{\pi}{m^2}\delta^3({\bf r})\right\}\nonumber \\
&+&2m(1-e^{-\beta mr}),
\end{eqnarray}
where $m$ is the effective mass of the gluon.

The first kind of singularity comes from the delta function which is
attractive for spin zero states. This singularity has been correctly
regularized by the authors with a standard method: the singular delta
function is replaced by a gaussian function (a Yukawa function is
another possibility)
\begin{equation}
\label{eq2}
\delta^3({\bf r})\rightarrow \frac{k^3m^3}{\pi^{3/2}}\, e^{-(kmr)^2}.
\end{equation}
Consequently, the $0^{++}$ ($L=S=0$) and $2^{++}$ ($L=0$, $S=2$) states,
for which the spin-orbit and the diagonal tensor interaction do not
contribute, have been correctly considered in the articles
\cite{hou03,hou01}. However, due to inaccurate numerical calculations,
the values of the masses obtained in Ref.~\cite{hou03} are questionable.
The values of the parameters found in this article are $\lambda=1.5$,
$\beta=0.5$ and $m\cong 670$ MeV (the gluon mass was not precisely
determined). We have calculated the masses of the
lightest $0^{++}$ and $2^{++}$ glueballs with the variational method
used in Ref.~\cite{hou03} and we have obtained an agreement with the
values published for $k=4.183$ (the value of this last parameter was not
given in the article). With these values of the four parameters, we have
performed an accurate calculation using two different numerical methods:
the Lagrange Mesh method \cite{bay86} and a variational method using an
harmonic oscillator basis up to 16 quanta. The masses of the $0^{++}$
and $2^{++}$ glueballs are found to be 282 MeV and 2238 MeV
respectively, which shows that the method used in Ref.~\cite{hou03} is
not accurate enough to reach convergence. Obviously, small variations of
the gluon mass cannot save the situation.

This defect cannot be totally compensated with another value of the
parameter $k$ (which carries less physical meaning). To reproduce the
masses of the $0^{++}$ and $2^{++}$ glueballs, both $k$ and $m$ must be
modified. With $m=705$ MeV and $k=3.298$, these masses are correctly
reproduced ($0^{++}$ mass is 1731 MeV and $2^{++}$ mass is 2402 MeV).
Even if the values of the parameters seem little changed, it
is worth noting that the wave functions, and consequently the associated
observables, are quite different. For instance, with these new
parameters, the root mean square radius of the lightest $0^{++}$
glueball is found to be 0.28 fm, whereas the published value is 0.1 fm.
Fortunately, the smearing of the delta function is only necessary for
the lightest scalar glueballs. Other states are not concerned by the
values of $k$.

Until now, we just pointed out some inaccuracies in the calculation,
which yields some difficulties in the interpretation of the parameters
and yields large errors for the value of the masses of glueballs
predicted by the models. Now we discuss more serious problems of the
model as treated in Ref.~\cite{hou03}.

The second kind of singularity of the potential (\ref{eq1}) appears in
the spin-orbit and in the tensor interactions (which were not considered
in the article \cite{hou01}). Indeed, these interactions feature an
unpleasant $r^{-3}$ singularity at the origin. Thus, when the spin-orbit
or the tensor terms, which comes from relativistic corrections, are
attractive the system collapses and has an
infinite negative binding energy. The standard approach to solve this
problem is to estimate the mass corrections by using the wave functions
computed without the relativistic corrections. This procedure was not
used in Ref.~\cite{hou03}. The authors prefer to treat
the singularity of the spin-orbit
and tensor terms by arbitrarily restricting the values of the
variational parameter $a$. Too large values are forbidden in order to
avoid the collapse. No real justification can be done for this
restriction. Consequently, the physical
meaning of most states presented in this article is questionable.

This last remark implies that the states which can be described without
any problems by the potential (\ref{eq1}) are $0^{++}$,
$2^{++}$ ($L=0$, $S=2$), $2^{++}$
($L=2$, $S=0$) (for these three states, the spin-orbit and the diagonal
tensor interactions are vanishing), $3^{++}$ ($L=2$, $S=2$) (the
spin-orbit interaction is zero while the diagonal tensor term is
repulsive), and $2^{-+}$ ($L=1$, $S=1$) (the repulsive spin-orbit
counteracts the attractive diagonal tensor interaction).

At last, from the expression (\ref{eq1}) for the interaction between two
gluons, it is easy to remark that the maximal value of the binding
energy is equal to $2m$, because the confining potential is
characterized by a saturation for $r\rightarrow \infty$. Thus the
maximal value for the mass of the glueball in this potential model is
equal to $4m$. Since the constituent gluon mass has been found to be (or
very close to) 670
MeV in Ref.~\cite{hou03}, the heaviest glueball has a mass equal to 2680
MeV. All the states having a mass greater than 2680 MeV are then
spurious.

To conclude, all the states presented by Hou and Wong are spurious or
characterized by an infinite negative binding energy, except three
states, the lightest $0^{++}$, $2^{++}$, $2^{-+}$.

F. Brau (FNRS Postdoctoral Researcher position) and C. Semay (FNRS
Research Associate position) would like to thank FNRS for
financial support.

\end{document}